\documentclass[aps,amsmath,amssymb,showpacs,showkeys,groupedaddress,endfloats,preprint]{revtex4}
\usepackage{dcolumn}
\usepackage{bm}
\usepackage[dvips]{graphicx}
\def \floatp@sw{\true@sw}
\makeatletter
\def\@dotsep{4.5}
\makeatother

\begin{document}

\title{Fitting of interatomic potentials without forces: a parallel particle swarm optimization algorithm}

\author{Diego Gonz\'alez}
\email{dgonzalez@gnm.cl}

\author{Sergio Davis}
\email{sdavis@gnm.cl}

% Afiliacion comun a todos los autores
\affiliation{Grupo de Nanomateriales, Departamento de F\'{i}sica, Facultad de
Ciencias, Universidad de Chile, Casilla 653, Santiago, Chile}

\date{\today}

\begin{abstract}
We present a methodology for fitting interatomic potentials to ab initio data,
using the particle swarm optimization (PSO) algorithm, needing only a set of
positions and energies as input. The prediction error of energies associated
with the fitted parameters can be close to 1 meV/atom or lower, for reference
energies having a standard deviation of about 0.5 eV/atom.
We tested our method by fitting a Sutton-Chen potential for copper from \emph{ab
initio} data, which is able to recover structural and dynamical properties, and
obtain a better agreement of the predicted melting point versus the experimental
value, as compared to the prediction of the standard Sutton-Chen parameters.
\end{abstract}


\maketitle

\section{Introduction}

In Condensed Matter Physics, the task of obtaining different mechanical properties of
materials, simulated atomistically with a large number of atoms under \emph{ab
initio} methods, is an almost prohibitive one, in terms of computational effort
with the current computer architectures. It might even at times be impossible. 
Because of this, producing a ``classical'' interatomic potential as a substitute
for the genuine quantum-mechanical interaction of the particles is highly
desirable. The usual procedure is to fit some empirical interatomic potential function, depending 
on $N$ parameters, requiring either agreement with certain macroscopic properties (structural,
thermodynamical, etc.) or simply agreement between the predicted and observed
energies and atomic forces. A standard algorithm based on force information is
the force matching method~\cite{Ercolessi1994, Izvekov2004}.

In this work we present a methodology for fitting interatomic potentials to
\emph{ab initio} data, using the particle swarm optimization (PSO)
algorithm~\cite{Kennedy1995}. The objective function to be minimized is the total prediction error in the
energies for the configurations provided, thus the algorithm does not require
any information besides the atomic positions for each configuration and their
corresponding \emph{ab initio} energies. In particular it does not require the atomic
forces, as in other fitting procedures such as force matching methods.

\section{Interatomic potential models}

We implemented two families of interatomic potentials, pair potentials and
embedded atom potentials. Among the former, we tested the well-known
Lennard-Jones potential~\cite{Allen87}, given by 

\begin{equation}
V(r) = 4\epsilon \Big[ \Big( \frac{\sigma}{r} \Big)^{12} - \Big(
\frac{\sigma}{r} \Big)^{6} \Big],
\end{equation}
and the 6-parameters ``generic'' potential as implemented in
Moldy~\cite{Refson2000},

\begin{equation}
V(r) = A\exp(-B r) + \frac{C}{r}^{12} -\frac{D}{r}^{4} - \frac{E}{r}^{6}
-\frac{F}{r}^{8}.
\end{equation}

From the family of embedded atom potentials~\cite{Daw1993}, having the general
form  

\begin{equation}
E_i = \frac{1}{2}\sum_{j\neq i} \phi(|\mathbf{r}_i - \mathbf{r}_j|) +
F\Big(\sum_{j\neq i} \psi(|\mathbf{r}_i-\mathbf{r}_j|)\Big).
\end{equation}
we implemented the Sutton-Chen potential, where the pair functions and the
embedding function are given by

\begin{eqnarray}
F(\rho) = \epsilon C\sqrt{\rho} \\
\phi(r) = \epsilon (a/r)^n \\
\psi(r) = (a/r)^m.
\end{eqnarray}

\section{Particle Swarm Optimization}

The particle swarm optimization (PSO) algorithm is based on the idea of
distributing the search procedure among a large number of ``agents'', which act
independently of each other. Each agent moves through the search space with a
simple dynamics, reacting to fictitious forces drawing it towards its own
\emph{current best} solution and the \emph{global best} solution for the whole
swarm. In this way, when an agent finds a better solution than the current
global best, it becomes the new global best and all the other agents react
instantly, the swarm is directed towards the new solution.

For a set of $n$ particles represented by their positions ${\mathbf{x}_1,
\mathbf{x}_2, ... , \mathbf{x}_n}$,
the velocity for the $i$-th particle and the $k$-th step is

\begin{equation}
\mathbf{v}_{i}^{k} = \omega \mathbf{v}_{i}^{k-1} + c_{1} r_{1}^{k} (\mathbf{x_B}
- \mathbf{x}_{i}^{k-1} ) + c_{2} r_{2}^{k} (\mathbf{x_G} - \mathbf{x}_{i}^{k-1} )
\end{equation}
and the position is given by

\begin{equation}
\mathbf{x}_{i}^{k} = \mathbf{x}_{i}^{k-1} + \mathbf{v}_{i}^{k}.
\end{equation}

We employed the following choice of PSO parameters: $\omega$=0.7, $c_1$=1.4 and
$c_2$=1.4, after a few trial convergence runs.

\section{Implementation of the fitting algorithm}

For a potential function where we wish to find the parameters $ {a_0, a_1, ... , a_m} $ 
from a set of positions $\mathbf{r}_i^j$ and energies $E_j$ satisfying the relation

\begin{equation}
V(\mathbf{r}_1^j, \mathbf{r}_2^j, \ldots,\mathbf{r}_n^j; \mathbf{a}) = E_j 
\end{equation}

with $$\mathbf{a} = (a_0, a_1, ... , a_m),$$

we can define an objective function which is just the total prediction squared
error, of the form

\begin{equation}
f(\mathbf{a}) = \sum_{j} \Big( V(\mathbf{r}_1^j, \mathbf{r}_2^j,
\ldots,\mathbf{r}_n^j; \mathbf{a}) - E_j \Big)^{2},
\end{equation}
and then for the set of parameters $\mathbf{a}^*$ that correctly fit the potential $V$ we have $f(\mathbf{a}^*) = 0$.

Then the problem may be solved numerically with the PSO algorithm minimising the function $f(\mathbf{a})$.

\subsection{Optimization of the algorithm}

We have included some improvements on the PSO implementation, particular to our
problem. For instance, we perturbed the swarm every time the procedure gets
stuck in a minimum for $N_S$ steps ($N_S$ proportional to the number of
parameters $d$ in the potential, usually $N_S=50d$), completely randomizing their positions. 

On the other hand, we exploit the fact that for several families of potentials
there is a scale parameter for the interatomic distance, let us call it
$\sigma$, such that the potential depends on $r$ only through $r/\sigma$. This
is the case for the $\sigma$ parameter in the Lennard-Jones potentials, for the
$C$, $D$, $E$, $F$ and $1/B$ parameters in the generic potential from Moldy, and also for the
$a$ parameter in the Sutton-Chen variant of the embedded atom potentials. 
This distance scale parameter can be constrained to be between the minimum
observed distance and a multiple of this value (typically 10 times), which 
considerably reduces the search space.

Parallelization was achieved simply by distributing the PSO particles evenly among the
different processors using the message passing interface (MPI) framework, at each
step sharing the global best between all processors.

\section{Results}

\subsection{Lennard-Jones potential}

In order to test the consistency of our procedure, we randomly generate~\cite{NoteConfigs} 
a set of 20 configurations and we compute their energy according to the standard Lennard-Jones parameters for argon,
$\epsilon$ = 0.0103048 eV and $\sigma$ = 3.41 \AA.

The resulting set has a standard deviation of energy of 0.41063 eV. Then, with the information 
of positions and energies (in a parallel run using 64 cores and 500 PSO particles), the time 
needed to find the minimum prediction error was 212.6 s. We can see that the algorithm converge quickly
for each parameter, recovering their exact values at 1300 steps (the prediction
error reached is below 10$^{-27}$ meV/atom).

\begin{figure}[h]
\begin{center}
\includegraphics[scale=0.6]{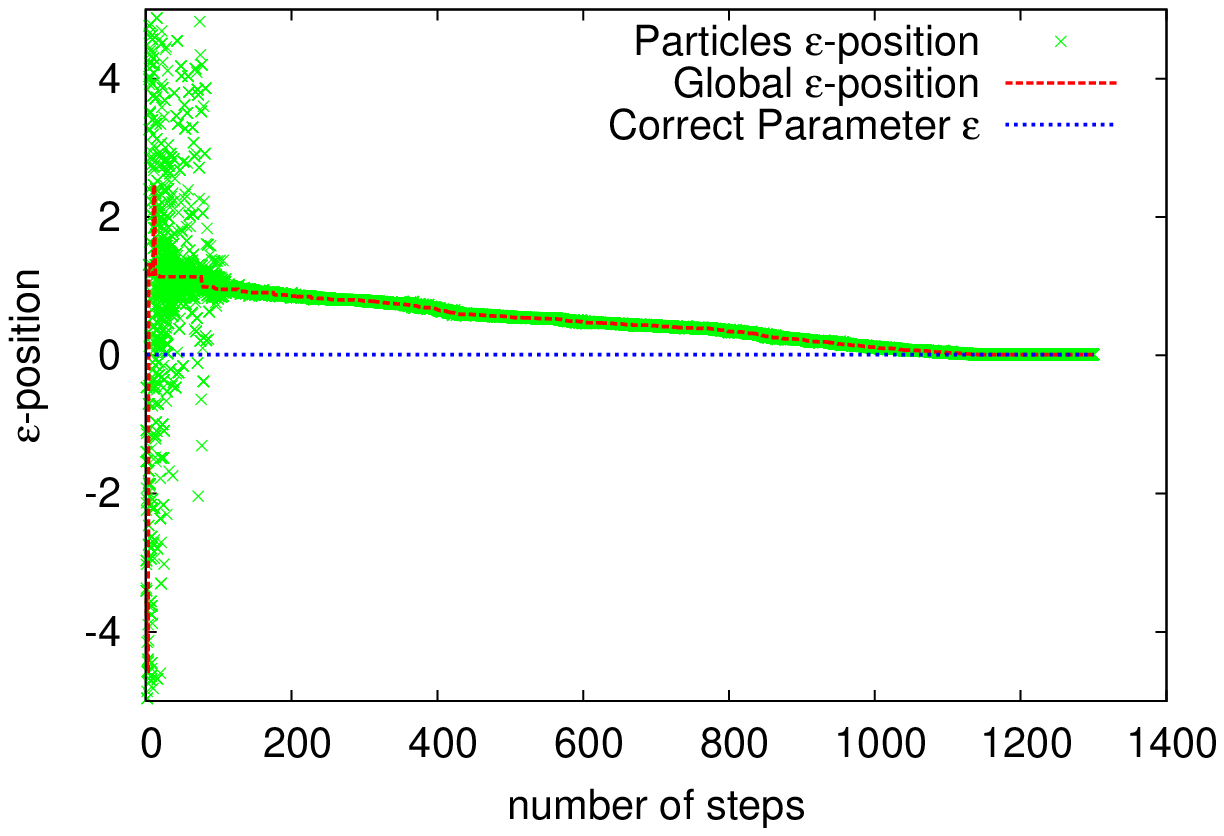}
\end{center}
\caption{(Color online) Evolution of the $\epsilon$ coordinate for the case of a Lennard-Jones
potential as a function of optimization step.}
\end{figure}

\begin{figure}[h]
\begin{center}
\includegraphics[scale=0.6]{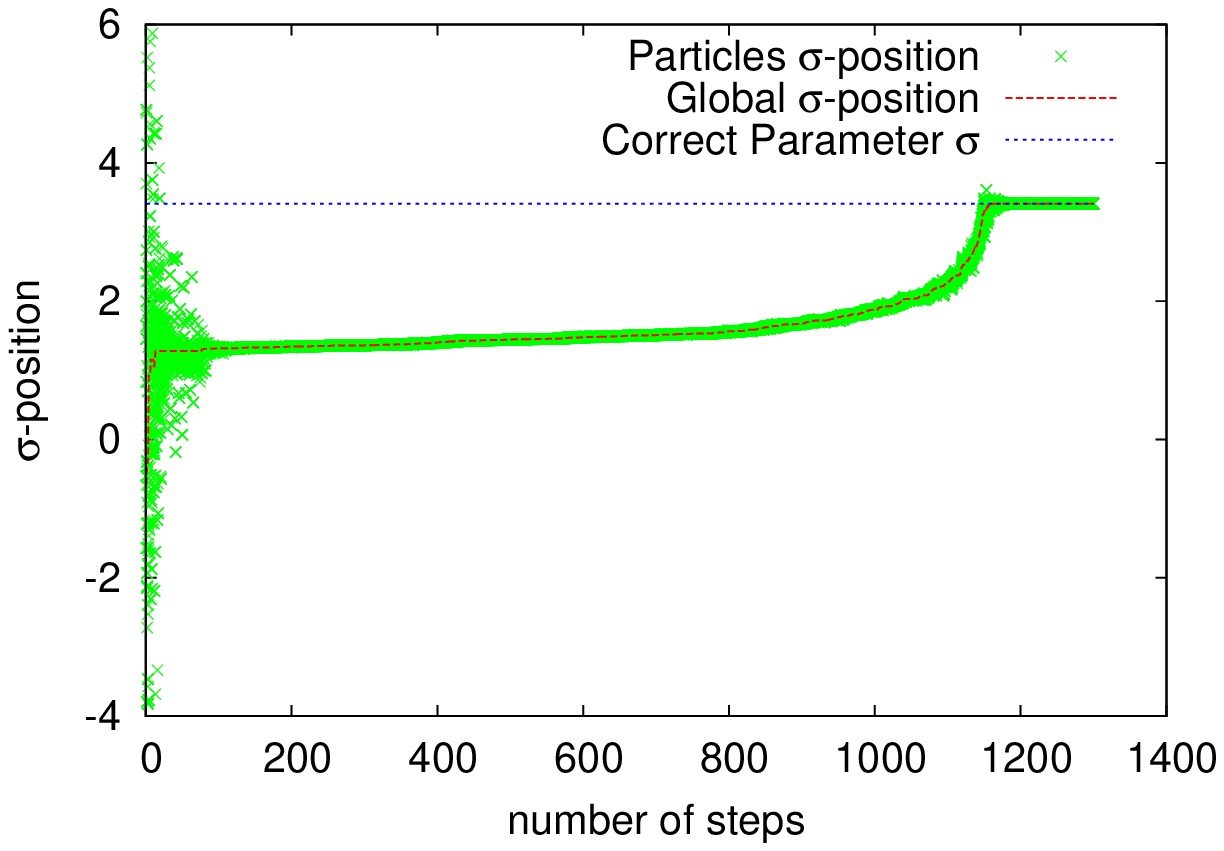}
\end{center}
\caption{(Color online) Evolution of the $\sigma$ coordinate for the case of a Lennard-Jones
potential as a function of optimization step.}
\end{figure}

\subsection{6-parameter generic potential}

For the 6-parameters pair potential using the same set of positions and
energies obtained for the previous Lennard-Jones test, the time needed to find
the minimum prediction error was 3159.9 s, again using 64 cores and 500
PSO particles. In this case the error for the converged set of parameters
falls below 8$\times$10$^{-2}$ meV/atom at 9000 steps.

\begin{figure}[h]
\begin{center}
\includegraphics[scale=0.65]{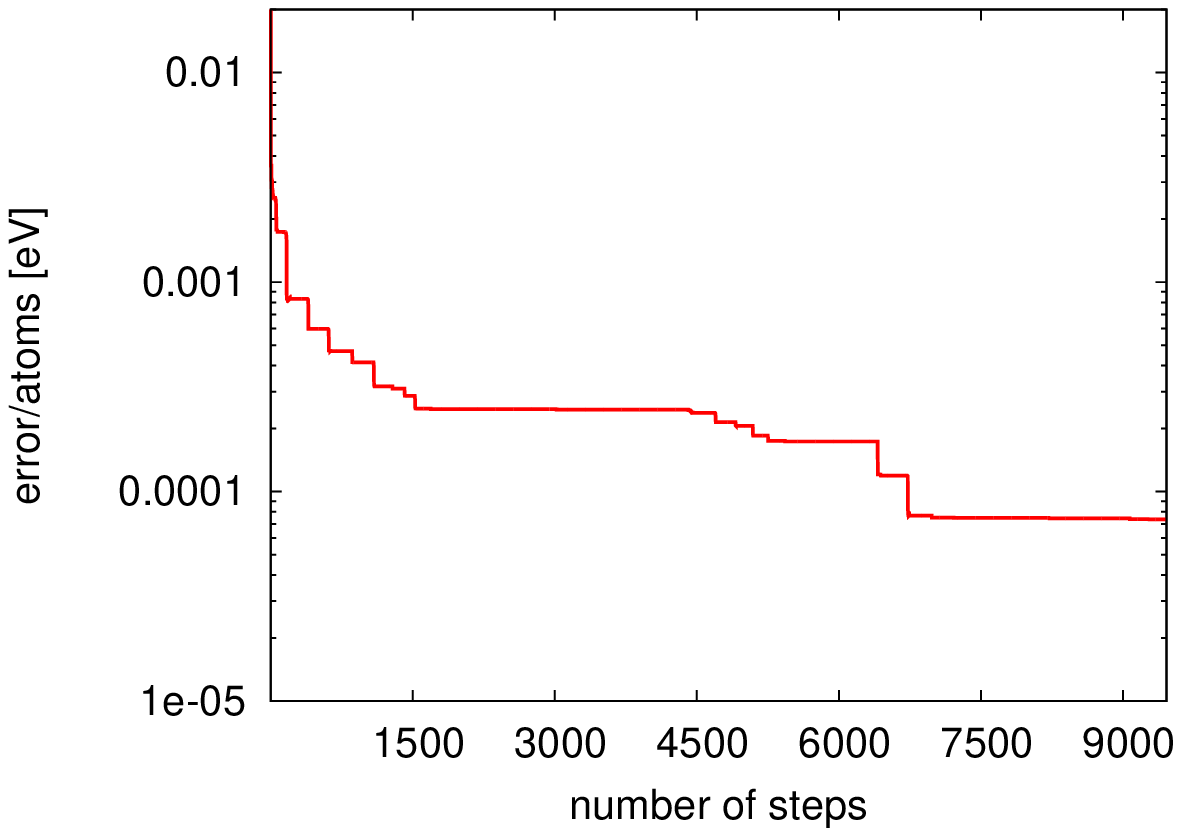}
\end{center}
\caption{Prediction error (meV) as a function of optimization steps for the case
of a 6-parameter pair potential.}
\end{figure}

\subsection{Embedded atom potential}

We repeated the same approach for the embedded atom potential, this time using
the standard Sutton-Chen parameters for copper, $\epsilon$=0.0123820 eV,
$a$=3.61 \AA, $n$=9, $m$=6 and $C$=39.432. We used 4 configurations as input,
and we stopped the minimization procedure after 193015 steps (execution time was
23 hours with 64 cores and 800 PSO particles), when we reached a prediction error 
of about 0.8 meV/atom and the following fitted parameters: $\epsilon$=0.0145749 eV, 
$a$=3.5834 \AA, $n$=8.82683, $m$=5.67465, and $C$=37.028.

\begin{figure}[h]
\begin{center}
\includegraphics[scale=0.65]{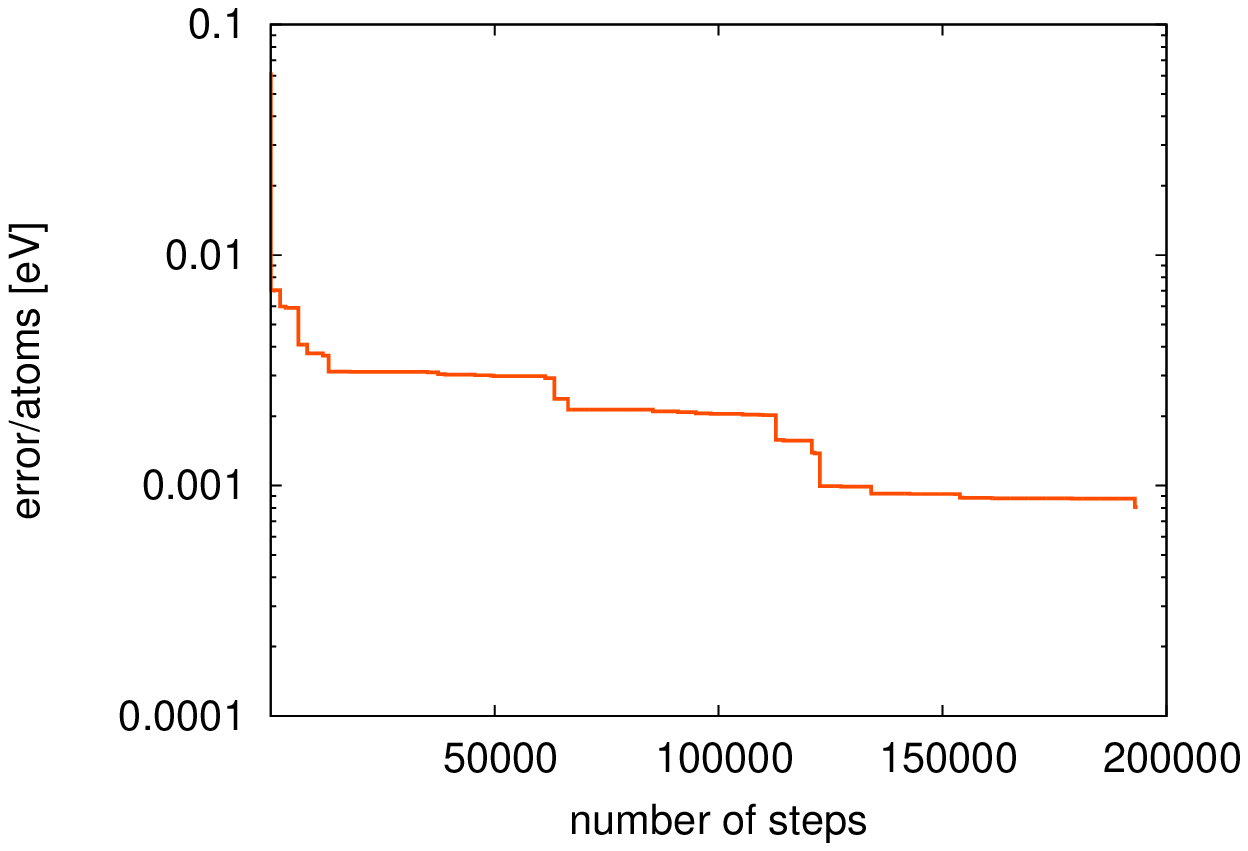}
\end{center}
\caption{Prediction error (meV) as a function of optimization steps for the case
of an embedded-atom potential.}
\end{figure}

\section{Application: an embedded atom potential for copper from \emph{ab initio} data}

In order to test our procedure on a more realist scenario and assess the quality
of the fitted potentials we performed \emph{ab initio} microcanonical molecular
dynamics simulations of copper at different temperatures (covering its solid, liquid and
superheated phases). All molecular dynamics calculations were performed using
Density Functional Theory (DFT) as implemented in VASP~\cite{Kresse1996}. We used
Perdew-Burke-Ernzerhof (PBE) generalized gradient approximation (GGA)
pseudopotentials~\cite{Perdew1996} with an energy cutoff of 204.9 eV and $k$-point
expansion around the $\Gamma$ point only.

From these simulations, we generated 13229 different atomic configurations with
their respective energies, mixed from solid, liquid and superheated state
simulations. Among them we chose a subset of 30 with maximum standard deviation
of the energy (namely 0.24 eV/atom), in order to increase the transferability of
the fitted potential. These configurations were used as input to the 
fitting procedure. We found the Sutton-Chen potential parameters presented in Table \ref{tbl_cu},
with a prediction error of 5.19 meV/atom.

\begin{table}
\begin{tabular}{|c|c|c|c|c|c|}
\hline
Source & $a$ (\AA) & $n$ & $m$ & $C$ & $\epsilon$ (eV) \\
\hline
Sutton and Chen & 3.61 & 9.0 & 6.0 & 39.432 & 0.012382 \\
\hline
Belonoshko \emph{et al} & 3.270 & 9.05 & 5.005 & 33.17 & 0.0225 \\
\hline
This work & 3.34385 & 5.93853 & 2.13419 & 32.2332 & 0.0846903 \\
\hline
\end{tabular}
\caption{Sutton-Chen potential parameters for Cu, fitted from ab initio data.}
\label{tbl_cu}
\end{table}

\begin{figure}[h]
\begin{center}
\includegraphics[scale=0.65]{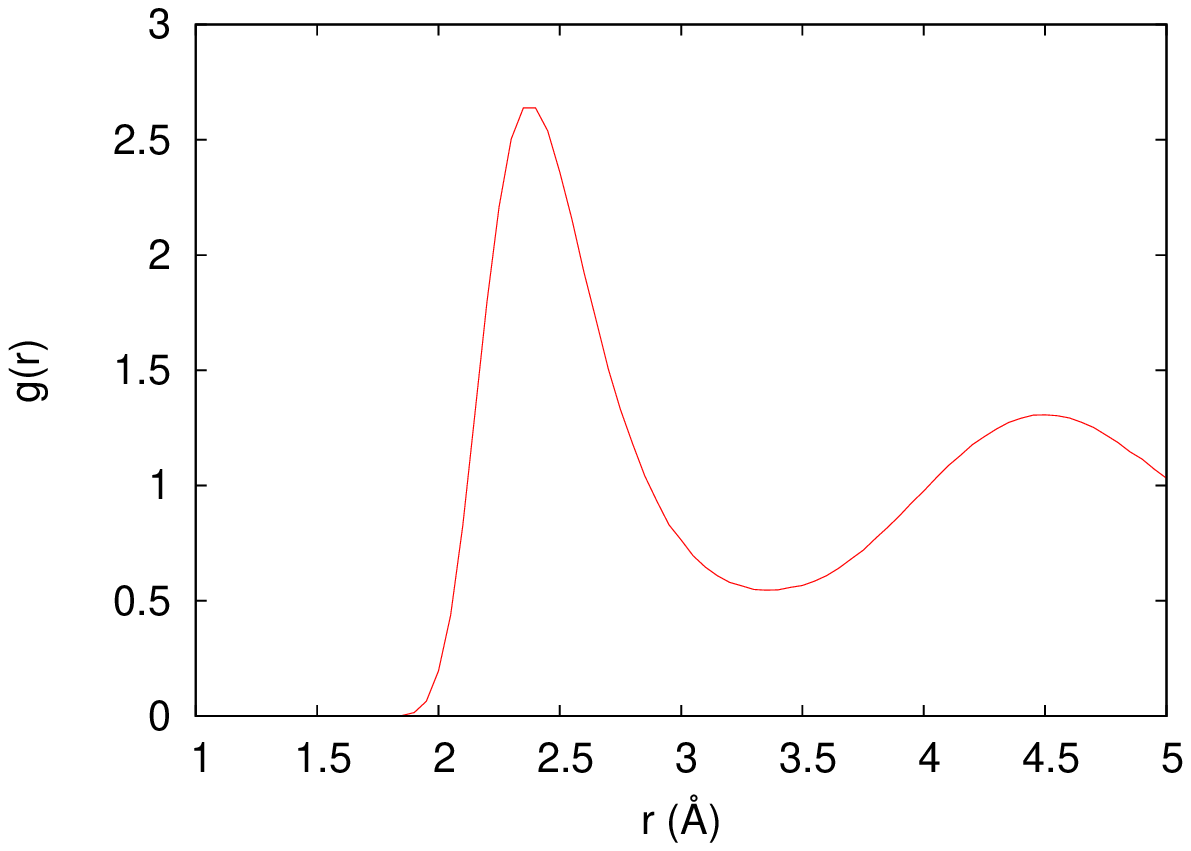}
\end{center}
\caption{Radial distribution function $g(r)$ for liquid copper at $T$=1500 K.}
\label{fig_cu_gdr}
\end{figure}

We tested these parameters by performing classical molecular dynamics
simulations using the LPMD~\cite{Davis2010} code, with a 4x4x4 FCC simulation cell
(256 atoms). Fig. \ref{fig_cu_gdr} shows the radial distribution function $g(r)$ produced by
our fitted Cu potential for liquid at $T$=1500 K. It reproduces exactly all features (positions of minima
and maxima, heights of the maxima) found in a previous ab initio fitting
study~\cite{Belonoshko2000}.

\begin{figure}[h]
\begin{center}
\includegraphics[scale=0.65]{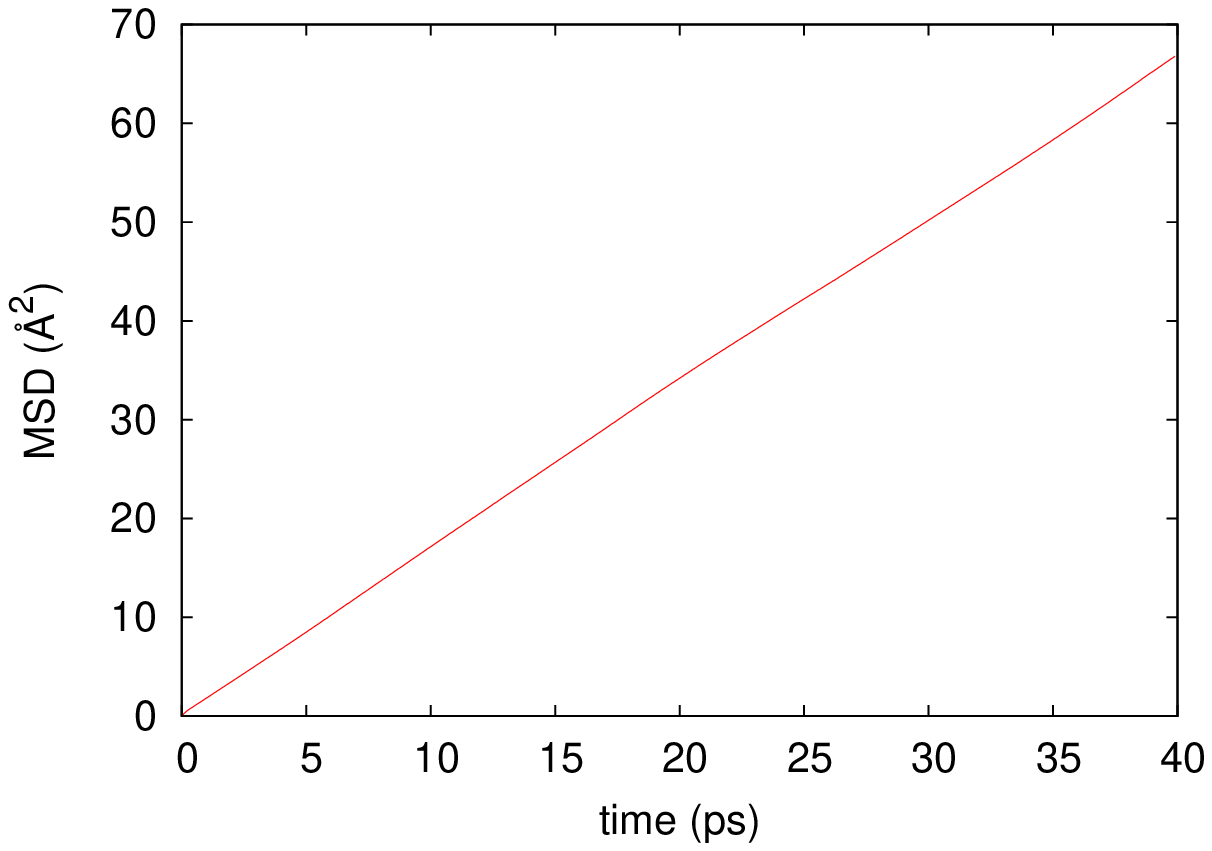}
\end{center}
\caption{Mean square displacement (MSD) for liquid copper at $T$=1500 K.}
\label{fig_cu_msd}
\end{figure}

Fig. \ref{fig_cu_msd} shows the mean square displacement for liquid at $T$=1500
K. From this we obtained a diffusion coefficient $D$=0.276924 \AA$^2$/ps, lower
than the experimental value reported by Meyer~\cite{Meyer2010}, 0.45 \AA$^2$/ps at $T$=1520 K.

The quality of the potential in reproducing thermal properties was assessed by
computing the melting point, using the microcanonical
Z-method~\cite{Belonoshko2006, Davis2010b, Davis2013}. 
In this method, for constant volume the $T(E)$ curve is drawn by performing
different molecular dynamics simulations at different initial kinetic energies
(in every simulation the system starts with the ideal crystalline
configuration). The discontinuity in the isochore signals the melting point.  

\begin{figure}[t]
\begin{center}
\includegraphics[scale=0.65]{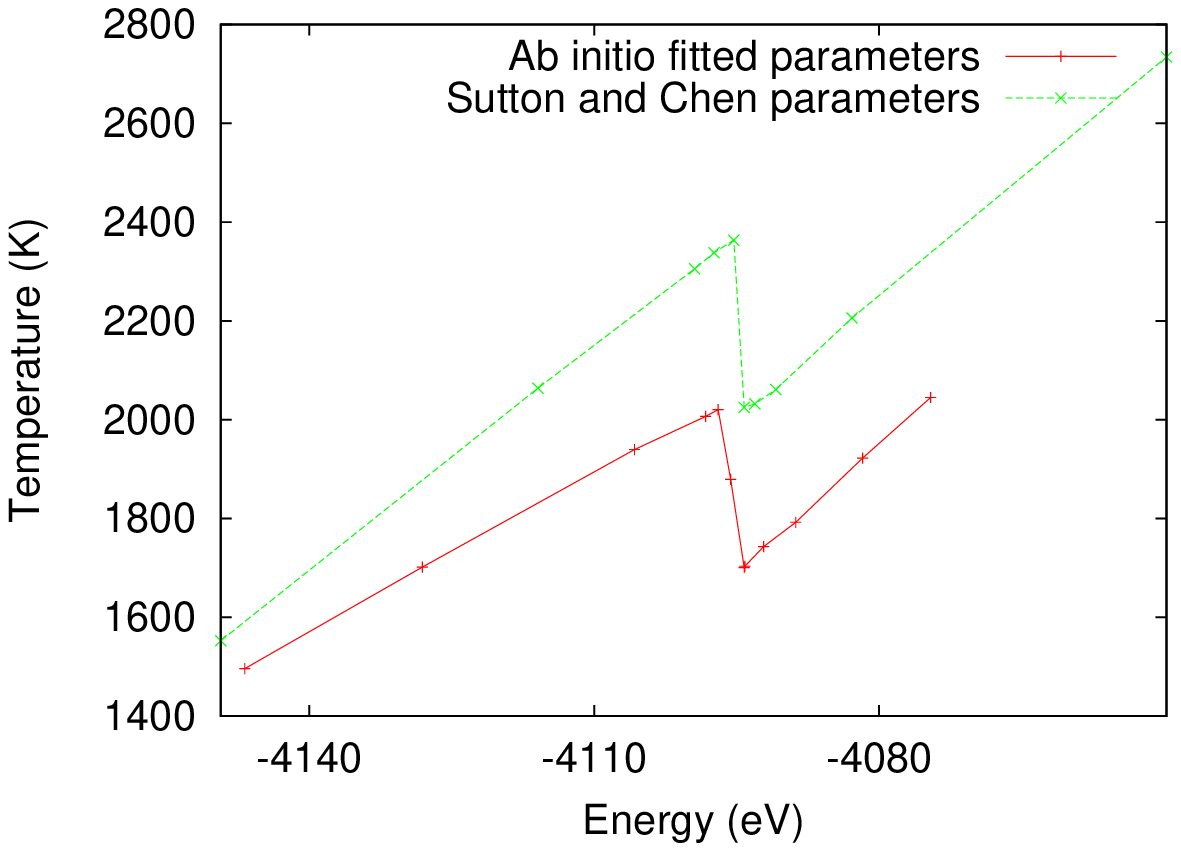}
\end{center}
\caption{(Color online) Isochoric curve (Z-curve) for copper according to our fitted potential
parameters. The Z-method predicts a melting temperature $T_m=$1700 K.}
\label{fig_cu_Z}
\end{figure}

Fig. \ref{fig_cu_Z} shows the isochoric curve for different energies around the
melting point, where the lowest point of the rightmost branch correspond to an
upper estimate of the melting temperature $T_m$, in our case approx. 1700 K (the
experimental value is $T_m$=1356.6 K). The highest point is the critical superheating 
temperature $T_{LS}$, around $T=$ 2020 K. For comparison we also included the 
isochoric curve calculated with the potential parameters by Sutton and Chen, 
which gives $T_m$ around 2000 K for the same system size and number of simulation steps.

\section{Concluding remarks}

We have shown that it is possible to use a parallel algorithm based on 
particle swarm optimization to fit interatomic potentials to \emph{ab initio} energies only.

Our procedure has been tested by fitting both pair potentials and embedded atom
potentials, up to a prediction error of the order of 1 meV/atom, using between 5
and 30 different configurations. The implementation code is parallelized using
message passing interface (MPI) libraries.

We demonstrated the capabilities of our method by fitting a set of Sutton-Chen parameters 
for copper using \emph{ab initio} data from three thermodynamic phases. This
fitted potential is able to reproduce the radial distribution function, although
it underestimates the diffusion coefficient for liquid copper at
$T$=1500 K (with respect to experimental data). It also yields a better prediction 
of the melting point than the standard Sutton-Chen parameters.

\begin{acknowledgments}
SD gratefully acknowledges funding from VID Universidad de Chile.
\end{acknowledgments}

%\bibliography{pso}

\begin{thebibliography}{15}
\bibitem[{Not()}]{NoteConfigs}
\bibinfo{note}{We generated random configurations by displacing the atoms of
  the ideal crystalline structure a small fraction of the lattice parameter.}
\end{thebibliography}
%\bibliographystyle{apsrev}

\end{document}